\documentstyle{aipproc}

\input epsf 

\begin{document}

\title{Data analysis techniques for stereo IACT systems}

\author{Werner~Hofmann, for the HEGRA collaboration}

\address{Max-Planck-Institut f\"ur Kernphysik, P.O. Box 103980,
        D-69029 Heidelberg, Germany}

\maketitle

\begin{abstract}
Based on data and Monte-Carlo simulations of the HEGRA
IACT system, improved analysis techniques were developed
for the determination of the shower geometry and shower energy from
the multiple Cherenkov images. These techniques allow, e.g., to select
subsamples of events with better than 3' angular resolution,
which are used to limit the rms radius of the VHE emission
region of the Crab Nebula to less than 1.5'. For gamma-rays
of the Mrk 501 data sample, the energy can be determined to typically
10\% and the core location to 2-3 m.  
\end{abstract}

Systems of imaging atmospheric Cherenkov telescopes (IACTs)
for TeV gamma-ray astronomy
allow the stereoscopic reconstruction of air showers, and
provide improved angular resolution, energy resolution, and
rejection of backgrounds such as showers induced by
cosmic rays, local muons, or random triggers caused by
night-sky background light. For systems with more than two
telescopes, the shower parameters are overdetermined, allowing
important cross-checks of the performance of the telescope
system and of the reconstruction algorithms. In particular,
the event-by-event determination of the position of the core
permits to directly measure effective detection areas, and to
estimate the systematic errors in the flux measurement
\cite{hofmann_kruger,mrk501}.

In this talk, I will cover recent developments concerning 
improved algorithms to reconstruct shower direction and shower
energy, and their tests using data from the HEGRA IACT system
\cite{performance,mrk501}. Detailed information as well
as a more complete list of references can
be found in \cite{recopaper,erespaper,crabsizepaper}.

\begin{figure}[htb]
\mbox{
\epsfxsize14.8cm
\epsffile{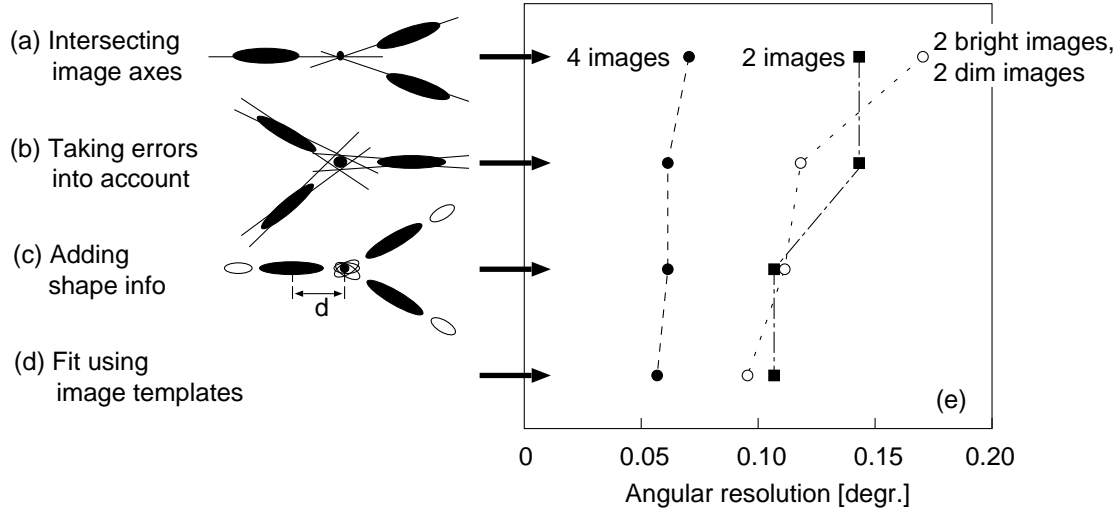}
}
\caption{
(left) Illustration of different algorithms to reconstruct
the shower direction from the multiple Cherenkov images.
(right) Resulting mean angular resolution for different data sets.}
\label{fig_reco}
\end{figure}

\begin{figure}[htb]
\mbox{
\epsfxsize14.5cm
\epsffile{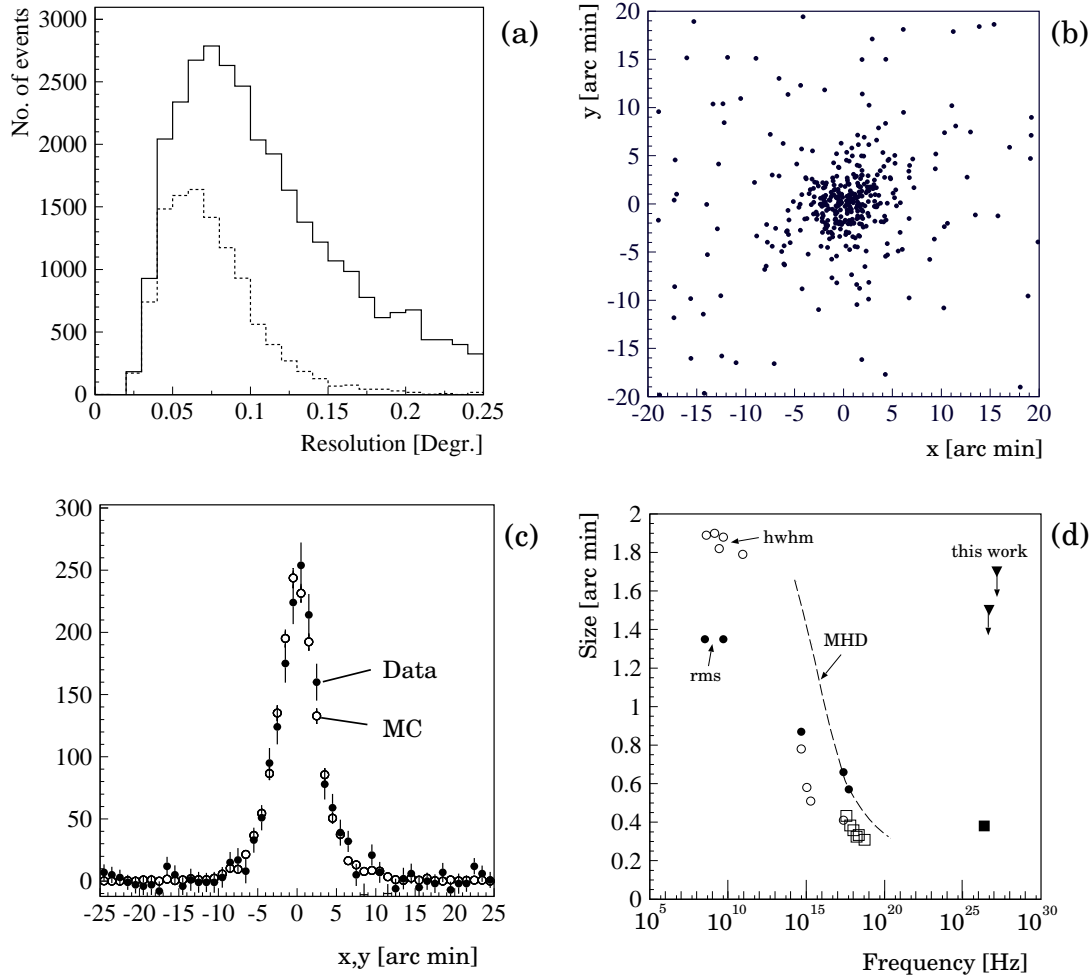}
}
\caption{
(a) Distribution of the estimated angular error for gamma-ray
events. The full line includes all events, the dashed line only
those where all four telescopes triggered.
(b) Angular distribution relative to the direction to
the Crab pulsar, for gamma-rays with estimated reconstruction errors
of less than 3' in each direction. (c) Distribution of gamma-ray
directions, compared to Monte-Carlo simulations assuming a point
source. (d) Rms radius of the photon emission region in the
Crab nebula, as a function of frequency, including the upper
limits obtained at TeV energies.}
\label{fig_dirres}
\end{figure}

{\bf Reconstruction of the shower geometry} \cite{recopaper}.
The traditional reconstruction algorithm used in HEGRA 
determines the shower direction by intersecting the axes
of all Cherenkov images, regardless of the quality of
individual images (Fig.~\ref{fig_reco}(a)). 
In particular in events combining some
bright images with dim images, the latter, with their
poorly determined image parameters, can spoil the
angular resolution. The angular resolution can be improved
by estimating, for each image, the errors on the image parameters
and by properly propagating these errors (Fig.~\ref{fig_reco}(b)).
In addition, one can use the shape of the image, in 
particular the {\em width/length} ratio, to estimate the 
{\em distance d} between the image centroid and the source, and use
this information to derive, for each telescope, an error ellipse
for the source location (actually, two ellipses, because of the
head-tail ambiguity). The ellipses from different telescopes are
then combined to locate the source (Fig.~\ref{fig_reco}(c)).
Finally, another approach (d) 
is to fit the intensity distribution
of the images using a set of image templates, rather than 
parameterizing images by their Hillas parameters.
Fig.~\ref{fig_reco}(e) shows the angular resolution for achieved
for different event classes. As expected, the techniques  (b)-(d)
outperform the simplest algorithm (a). The fit (d) is generally
best, but the improvement compared to the much simpler and faster algorithm
(c) is not dramatic. In addition to an improvement in the 
angular resolution, algorithms (b)-(d) provide, for each event,
an estimate of the angular resolution (Fig.~\ref{fig_dirres}(a)),
which can be used, e.g., to reject poorly reconstructed events.

{\bf Size of VHE emission region of the Crab Nebula} \cite{crabsizepaper}.
Well-reconstructed events reach an angular resolution on the same
scale as the characteristic size of the Crab Nebula. One can use
such events to search for evidence for an extended VHE emission 
region. Fig.~\ref{fig_dirres}(b) shows the angular distribution
of events with an estimated angular error of less than 3' in 
each projection, relative to the direction to the Crab. The 
width of the distribution is, within statistical errors, identical
with the width expected for a point source
on the basis of simulations (Fig.~\ref{fig_dirres}(c))
and with the width of the gamma-ray distribution observed for 
Mrk 501. Therefore, we can only give an upper limit on the size
of the emission region. Including systematic effects, e.g. due to
pointing errors, we find a 99\% upper limit on the rms radius
$<r^2>$ of the TeV emission region of 1.5'. This value is comparable
to the radius at radio wavelengths, but significantly larger than
the size at x-ray energies. Standard models for the VHE gamma-ray emission
of the Crab Nebula assume that the same electron population is
responsible for x-rays via synchrotron emission, and for TeV gamma-rays
via the IC process, and predict a small TeV emission region, well below
 the experimental limit. 
Possible hadronic production of gamma-rays, on the other hand, could
take place at significantly larger distances from the pulsar.

{\bf Core determination} \cite{erespaper}. The shower core is usually located
by intersecting the image axes, starting from the telescope locations.
The precision of the core determination is therefore given by
the precision with which the image axes can be determined, typically
$O(5^\circ)$. If the source location is known, as is the case, e.g.,
for the Mrk 501 data sample, one can alternatively determine the
image axis as the line connecting the image of the source and the
image centroid. With a typical distance between the source and the
image centroid of $1^\circ$ and a measurement of the centroid to
$O(0.02^\circ)$, the image axis is then known to $O(1^\circ)$.
Using this technique, Monte Carlo simulations predict that the 
precision for the shower core improves from about 6~m to 10~m
for the normal method, to about 2~m to 3~m, depending on the 
core distance (in each case, properly taking into
account the errors on the measured image parameters). 
The exact knowledge of the core position is particularly
important for the energy determination, when the observed light 
yield is translated into an energy estimate.

\begin{figure}[htb]
\mbox{
\epsfxsize14.5cm
\epsffile{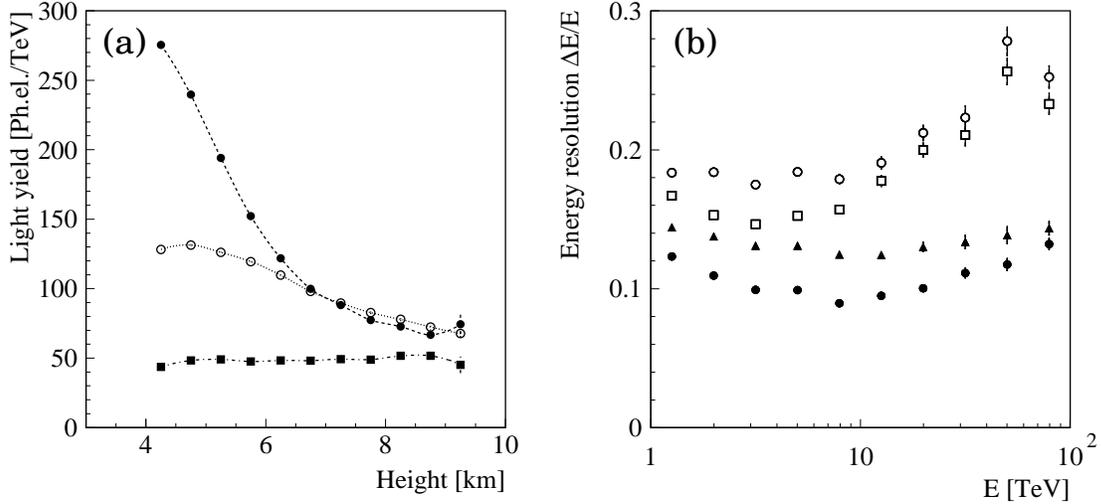}
}
\caption{
(a) Light yield (in Photoelectrons/TeV) as a function of the height
of maximum Cherenkov emission, at core distances around 40-50~m (full circles),
90-100~m (open circles) and 140-150~m (full squares). (b) Energy resolution
as a function of energy, for the conventional energy reconstruction
(open circles), with improved core determination (open squares),
with shower height correction (full triangles), and with 
shower height correction and improved core determination (full
circles).
}
\label{fig_energy}
\end{figure}

{\bf Energy determination} \cite{erespaper}. 
Earlier studies comparing event-by-event
the light yield observed in different telescopes indicated correlated
fluctuations in the light yield of individual showers
\cite{hofmann_kruger}. Monte-Carlo studies point to the fluctuation
in the height of the shower maximum as the primary source for these
correlated fluctuations. Fig.~\ref{fig_energy}(a) illustrates that
for distances up to about 100~m from the shower axis, the light yield
varies significantly with the height of the shower; only beyond the
Cherenkov radius of about 120~m is the light yield stable. An obvious
approach to improve the energy resolution is therefore to measure the
height $h_{max}$ of the shower maximum, 
and to include it as an additional parameter,
writing $E_i = f(size_i,r_i,h_{max})$, where $size_i$ is the image
{\em size} measured in telescope $i$ at a distance $r_i$ from the
shower axis. With an IACT system, the height of the shower maximum, 
or, more precisely, the height of maximum Cherenkov emission, can 
be determined essentially by triangulation, using the relation
between the {\em distance} $d_i$ from the image to the source,
$r_i$, and $h_{max}$: $d_i \approx r_i/h_{max}$. The actual
algorithm \cite{erespaper} uses a slightly more complicated
relation, reflecting the fact that light arriving at small $r_i$
is generated by the tail of the shower rather than by particles near
the shower maximum. The algorithm reaches a resolution in shower
height of 530 to 600~m rms.

Fig.~\ref{fig_energy}(b) illustrates the effect of the various
possible improvements to the energy resolution. Whereas the
conventional algorithm provides a resolution of 18\% to 22\%
for the 1~TeV to 30~TeV range, the shower-height correction
provides a resolution of about 12\% to 14\%, and the combination of the
shower height correction with the improved core determination
assuming a known source yields 9\% to 12\% resolution.

Before applying this technique to the actual data to obtain
improved energy spectra, one needs to make sure that systematic
effects are under control at a level consistent with the 
improved resolution. While the redundant data from the IACT 
system provide sufficient information to check this, the analysis
is not yet finished. A first test of the new method with Mrk 501 data
results in a spectrum consistent with earlier analyses, possibly
with a slightly steeper spectrum in the cutoff region beyond 6 TeV.

{\bf Summary.} The analysis algorithms discussed here represent
clear improvements over the first-generation algorithms used in
the reconstruction of data from the HEGRA IACT system; it is also
clear that further improvements are possible and that at this
point we do not fully use all the information provided by multiple
IACT images of an air shower. The algorithms do not only improve
the angular resolution and the energy resolution; they also help
to boost the significance of faint signals. For example, instead
of simply counting all events reconstructed 
within a certain angular distance form a
source, one can form a weighted sum, weighting events according
to their expected signal-to-background ratio, as determined 
event-by-event from the estimated angular error and misidentification
probability. First tests of such methods indicate in an increase
in the significance for the detection of weak sources by up to 80\%.

{\bf Acknowledgments.}
Many of the members of the MPIK CT group have contributed in one
way or another to
the development and tests of the advanced analysis techniques discussed here;
in particular, I. Jung, A. Konopelko, H. Lampeitl, H. Krawczynski and 
G. P\"uhlhofer should be mentioned.

\end{document}